\documentclass[12pt]{article}
\usepackage{amsmath,amssymb,mathrsfs}
\usepackage{graphics}
\usepackage{epsfig}
\usepackage{graphicx}
\usepackage{float}

\usepackage{color}
\usepackage{setspace}

\newtheorem{thm}{Theorem}[section]

\newtheorem{rema}[thm]{Remark}

\numberwithin{equation}{section}
\def\theequation{\thesection.\arabic{equation}}
\def\be{\begin{eqnarray}}
\def\ee{\end{eqnarray}}
\def\numero{\refstepcounter{equation} (\theequation)}

\let\text=\textstyle

\def\be{\begin{eqnarray}}
\def\ee{\end{eqnarray}}
\def\ben{\begin{eqnarray*}}
\def\een{\end{eqnarray*}}
\def\bei{\begin{itemize}}
\def\eei{\end{itemize}}

\title{\bf  Interception efficiency of CVM-based lightning protection systems for  buildings and the fractional Poisson model}

\author{Harold S. Haller$^{1}$ and Wojbor A. Woyczynski$^{2}$}

\date{\today}

\begin{document}

\maketitle

\footnotetext[1]{Director, Case Statistical Consulting Center, Department of Mathematics, Applied Mathematics, and Statistics, Case Western Reserve University }

\footnotetext[2]{Professor, Department of Mathematics, Applied Mathematics and Statistics, and Director, Center for Stochastic and Chaotic Processes in Science and Technology, Case Western Reserve University, Cleveland, OH 44106, U.S.A., email: waw@case.edu, http:/sites.google.com/a/case.edu/waw}



  {\bf Abstract:}  The purpose of this paper is to resolve a question regarding efficiency of a lightning protection system (LPS) for buildings based on the collection volume method (CVM) . The paper has two components. The first, following suggestions of other authors [Abidin and Ibrahim 2004],  takes advantage of count data from installed devices,  and independent installation-site inspections  to develop  our statistical analysis. The second component  investigates the validity of the underlying theory by introducing a novel   methodology of fractional Poisson processes, which are able to  reproduce the burstiness of lightning strikes,   an essential feature   of stochastic   time dependence  of incidence   of lightning strikes.   The   standard Poisson processes   used in the past efforts in this area cannot do that.

\bigskip



\section{Introduction}

  This paper provides the statistical analysis of an unprecedented field study for the assessment of interception efficiency of a  lightning protection system (LPS) based on optimally positioned    air terminals with the optimality assured by the Collection Volume Method (CVM).  It also compares these data with theory using extensions and enhancements to theoretical models of the equivalent exposure area ($A_{eq}$) and attractive radius ($R_a$) [D'Alessandro and Petrov 2006].

  Between 2010 and 2012, buildings protected by a system of  air terminals optimally placed according to the CVM lightning protection methodology\footnote{Designed and build by ERICO$ ^{\footnotesize\textregistered}$.}  were surveyed in Kuala Lumpur, Malaysia, by T\"UV-Hessen, an independent expert organization based in Germany. At each installation, T\"UV-Hessen surveyed the buildings and documented evidence of lightning damage (bypasses) and recorded the readings of instruments showing the number of captured lightning events. At the end of the third round of inspections, 33 events had been collected over a combined 37 terminal-years of exposure. This agrees closely with the expected number of events determined by the Eriksson $R_a$ model [D'Alessandro and Gumley 2001].  Furthermore, the results are in agreement with a field study carried out earlier in Hong Kong, which confirms the relevance of the Eriksson model [Petrov and D'Alessandro 2002]. Finally, the average interception efficiency of the lightning protection systems was measured against the predicted average interception efficiency on which the CVM-optimized terminal placement had been based. The average interception efficiency was found to be in very close agreement with the predicted efficiency. This confirms the result of a previous field study in Kuala Lumpur, the results of which were published in  [D'Alessandro and Petrov  2006].

  At the fundamental level the paper makes an effort to provide in this field a novel  mathematical model that would be able to reproduce the random burstiness of lightning strikes. Burstiness is commonly observable in many time-dependent  phenomena, such as natural disasters,  network/data/email network, or vehicular traffic. It is, in part, due to changes in the probability distribution of inter-event times: distributions of bursty processes or events are characterized by heavy, or fat, probability tails, and with this  observation in mind we have applied in this paper the concept of a fractional Poisson process (fPp)  [Cahoy, Uchaykin, and Woyczynski 2010]\footnote{Another approach to burstiness is based on the so-called  Fano factorÑa ratio between the variance and mean of counts. For other approaches to the  burstiness problem for  point processes, see [Neuts 1993].}  The   standard Poisson processes   used in the past efforts in this area cannot accommodate the burstiness, see, e.g., [Petrov and DÕAlessandro 2002], which accepts the Poisson hypothesis based on a Chi-square value with only three degrees of freedom.
 Also, in the case of our data,  the fit  via the fractional Poisson process has been demonstrated to be superior to the standard Poisson model used previously in this field, but we recognize that to further support the claim of robustness of the fPp comparisons with other lightning strike data are needed.

It is essential  to acknowledge that lightning is a stochastic natural event and there are no lightning models that are 100$\%$ accurate. Similarly, there are no known lightning protection systems that are 100$\%$ efficient. For this reason, field testing methods as reported in this paper are particularly relevant. Furthermore, it is important to note that the reader should pay particular attention to the documents listed in the bibliography because they provide a lens by which to view the controversy concerning conventional and unconventional LPS.
  	In particular, a paper published some ten years ago [Abidin and Ibrahim 2004] referred to the absence of reliable evidence that CVM air terminals offer an increased zone of protection over the conventional air terminal (a.k.a. Franklin rod). The independently assessed lightning strike data from buildings in Kuala Lumpur, Malaysia, with optimally installed CVM air terminals that we reviewed provides one source of reliable evidence of the efficiency of these CVM systems. However, no data were available for our review from which a contrast could be made between optimally positioned Franklin and CVM air terminals. Only a statistically designed experiment, which controls for building size and location biases, will resolve this conflict.

	The results detailed in this paper rely heavily on the prior work in the field of lightning protection.  In particular, see [Anderson and Eriksson 1980, Petrov and DÕAlessandro 2002]. The theory of LPS in general, and CVM models in particular, which are based on an electrogeometric model (EGM) of striking distance and peak stroke current, are partially empirical. As such, when field studies are compared with any of these theories, both the pure physics and the empirical aspects of the models are evaluated. From a statistical point of view, our intent is to determine if there is a statistically significant difference between theory and data based on the uncertainty in the data.  
	
	For a general exposition on a variety of stochastic models in geosystems see [Molchanov and Woyczynski 1997] and, in particular, [Klyatskin and Woyczynski 1997]. 


  \section{Preliminaries}

  
 \subsection{Scope of the paper}
The efficiency of a lightning protection system (LPS) depends on the placement of lightning rods on the structure to be protected as well as the design of the lightning rod or air terminal. These lightning rods can be placed according to various models currently used in the field of lightning protection, all of which are based on the physical properties of lightning, and many years of observations. It is important that these models be scientifically verified in-situ in order to assess their effectiveness with respect to the standardized lightning protection level (LPL). One such standard for lightning protection is presented in the IEC 62305 series of standards. This paper pertains to the documentation of a field trial to verify the validity of the Collection Volume Method (CVM) model. Publications of prior field validations include a Hong Kong study [Petrov and D'Alessandro 2002],  and a previous study in Kuala Lumpur [D'Alessandro and Petrov 2006]. While such unique field studies have been conducted for this model and published in scientific journals, it was determined that a continuation of the Kuala Lumpur field study would be beneficial to further validate the CVM model. This is particularly important in light of articles that claim there is no reliable evidence that the un-conventional lightning air terminals offer an increased zone of protection over that of conventional terminals (a.k.a. Franklin lightning rods) [Abidin and Ibrahim 2004].

	In order to properly execute a continuation of the prior studies, collaboration was established with an independent technical agency, T\"UV-Hessen,  with expertise in safety assessment. The independent firm had the responsibility of collecting lightning data as detailed in Section 3. The scope of this paper is to analyze the collected data using the main steps of the statistical method employed  in [D'Alessandro and Petrov 2006] in order to determine the interception efficiency of CVM-based lightning protection systems, and to compare these results using extensions and enhancements to theoretical models of the equivalent exposure area ($A_{eq}$) and attractive radius ($R_a$) (D'Alessandro and Petrov 2006).
 

\subsection{CVM-based lightning protection system}
The lightning protection systems surveyed during this study are known as ERICO$ ^{\footnotesize\textregistered}$  SYSTEM 3000  lightning protection systems (LPS). A typical LPS is comprised of an optimized air terminal that is grounded using an insulated downconductor. A lightning event counter was included in each LPS to count the current impulses from lightning collected by the air terminal.\footnote{The systems were installed according  to  the  ERICO$^{\footnotesize\textregistered}$    Installation,  Operation  and  Maintenance manual [1]. ERICO is a registered trademark of ERICO International Corporation.} Thus a statistical evaluation of a LPS relates to a combination of air terminal design and positioning of the air terminals on each building.

	For each site, the air terminals were placed according to the Collection Volume Method, which is extensively described in [D'Alessandro and Gumley 2001] and [D'Alessandro 2003]. More recently, the CVM has been cited in the 2012 edition of the IEEE Guide for Direct Lightning Stroke Shielding of Substations, [IEEE998-2012]. As explained in [D'Alessandro  and  Gumley  2001],  the  CVM  is  based on the Eriksson attractive radius model. According to [IEC 62305-1:2010], lightning current is a parameter used to calculate the radius of protection in a lightning protection system. Each level of lightning protection is based on a minimum lightning current value.
	
	A proprietary software package, LPSD 3.0, offers a method to implement 3D models of structures and the placement of air terminals using the CVM. The fundamentals of LPSD are explained in [D'Alessandro and Gumley 2001], and [D'Alessandro 2003].  The software was used in the above described  T\"UV study to model each lightning protection design, and determine the location of the optimized air terminals according to a specified level of protection.
 
 \begin{figure}[ht]
\centering
\numberwithin{figure}{section}
\includegraphics[width=8cm]{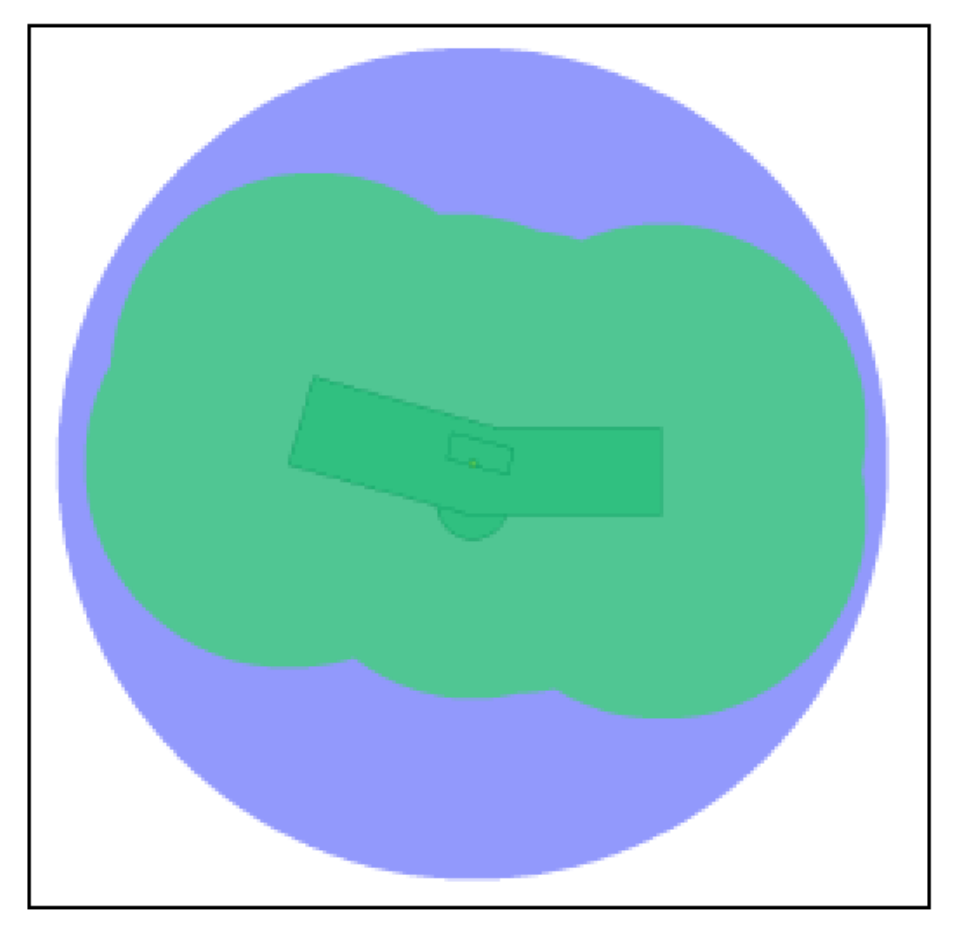}
\caption{Placement of an optimized air terminal in LPSD 3.0 according to the CVM.}
 \end{figure}
 
 \newpage

	Figure 2.1 shows an LPSD 3.0 screenshot of the plan view of a building protected by optimized air terminals according to the CVM. This particular building was surveyed in the T\"UV  study. The blue color disk  represents the area of coverage of the optimized air terminals while the green area shows the competing features of the building. The output image shows that the design meets the specified protection level when the blue area (i.e. the area of coverage of the optimized air terminal) fully encloses the green area (i.e. the competing features of the structure). The importance of Figure 2.1 lies in the fact that the field study for our report reflects data from CVM air terminals that have been optimally positioned on the buildings to provide the coverage indicated by blue in Figure 2.1. Consequently the statistical analysis of these data reflects both CVM air terminals and optimal positioning of the devices. Damage to buildings noted in [Abidin and Ibrahim 2002] used single Early Streamer Emission (ESE) air terminals, which may or may not have been optimally positioned. 
 

\subsection{ Review of past field studies}
The two real-world field studies referenced in this report, namely [Petrov and D'Ale\-ssandro 2002], and [D'Alessandro and Petrov 2006], were published in the Proceedings of the Royal Society of London A. The studies were conducted in Hong Kong and Kuala Lumpur, respectively. Both of these areas are known for their high keraunic levels and are therefore excellent locations for long-term field studies of lightning protection systems.
  
  \smallskip

In the {\it 2002 Petrov and  D'Alessandro study}, various models of lightning interception were analyzed against real lightning strike data collected in Hong Kong. The field data spanned a period of 8 years and were taken from 161 structures ranging in height from 10 m to 370 m. These structures were protected with lightning air terminals that had been positioned in the optimal roof locations using an improved electrogeometric model, namely the    CVM.

	One of the goals of this research was to determine the validity of Eriksson's attractive radius model. In doing so, the effectiveness of the CVM was assessed.   Table 1 in [Petrov and D'Alessandro 2002] shows the mean value of the total expected number of strikes for the attractive radius models considered to be 111,  while the total observed number of flashes was 103.  Therefore, approximately 92$\%$ of all flashes are estimated to have been intercepted by the lightning protection systems. This value is consistent with the typical design interception efficiency expected from traditional lightning protection systems based on the Rolling Sphere Method (RSM), or other similar approaches.
	
	In summary, the Hong Kong study demonstrated the validity of the attractive radius models, upon which the CVM is based, by showing that the striking distance is a function of the height of the structure. Also, using the overall statistics, an interception probability of around 90$\%$ was estimated.

  \smallskip

The {\it 2006 D'Alessandro and  Petrov study}  included the analysis of 13 years of lightning strike and bypass data for buildings in the Klang Valley/Kuala Lumpur region, one of the most active lightning regions of the world.  The data sample was comprised of 86 structures with a mean height of 57 m and mean exposure time of 6.9 years. These structures were subjected to a combined 384 flashes over a total combined observation time of 592 years.  Each building was  equipped with a lightning protection system.

	Since an LPS with 100$\%$ efficiency does not exist, the major aim of this study was to determine the proportion of strikes captured by the LPS out of the total number of incident strikes. This quantitative study was remarkable in the sense that such an analysis had not been published since the origins of lightning science more than 250 years ago. After the application of a wide range of statistical tests on the data, it was found that the percentage of strikes captured was about 87$\%$. This is in agreement with three leading theoretical models of lightning interception. Using a known median current of 33 kA for Malaysia, the theoretical interception attainable from the Eriksson and Petrov models was 86$\%$ and 83$\%$ respectively.  Hence, both were in excellent agreement with the observed protection level of 87.5$\%$.


\section{Data Collection Review}

\subsection{T\"UV-Hessen third-party inspections}
The data collection in this study was carried out in the Klang Valley region of Malaysia over the period 2010-2012. For each building surveyed during each round of yearly inspections, the T\"UV-Hessen inspector prepared an inspection report containing the following information:

$\bullet$	Name of the building;

$\bullet$	Date and time of inspection;

$\bullet$	Numbers of the corresponding pictures taken of each LPS installation and of the eventual bypasses on each building;

$\bullet$	Information about the building (roof and wall material);

$\bullet$	General information about the LPS (number of terminals, location of terminals against LPSD 3.0 design report);

$\bullet$	Condition of the LPS (optimized terminal and downconductor condition, resistance of grounding system);

$\bullet$	Result of lightning damage due to bypasses investigation on roof and upper part of the walls (including detailed sketch of building roof);

$\bullet$	Lightning counter reading.

 \medskip

Following each round of inspections in 2010, 2011 and 2012, a final report was compiled by T\"UV- Hessen that contained:

$\bullet$	Information about the scope of the study, approach followed to carry out the study and T\"UV-Hessen inspector credentials;

$\bullet$	General information about the inspection process and inspection document;

$\bullet$	A summary of inspections results in tabular format.

\medskip
  
The present study is based on data collected from lightning event counters on the lightning protection systems of 17 buildings. The initial inspection in 2010 was performed on all 17 buildings with 6 of them being part of the previous Kuala Lumpur study.  However, data were inaccessible on four of these buildings.  Hence 2011 was considered as being Óyear zeroÓ for these four buildings instead of 2010 for the 13 others. The summary of the T\"UV-Hessen 2010 final report indicated that the approach taken, and the results,  were satisfactory and that inspections could be continued in 2011. In 2011, after one year of exposure, these 17 buildings were re-inspected. In order to collect more data more rapidly, 16 other buildings were added to the study,  increasing the size  of  the building  sample  to  33.  A full  Óyear zeroÓ  inspection was carried out on 12 of the 16 newly added   buildings.

One newly added site was considered by the inspector to be at Óyear 1Ó due to the fact that the counter had been installed six months earlier and had already recorded lightning flashes. This assumption is conservative from a statistical standpoint.

\newpage

Four other sites were not inspected in 2012 for one or more of the following   reasons:

$\bullet$	Grounding electrode not accessible for inspection;

$\bullet$		Roof not accessible;

$\bullet$		Downconductor routing not per manufacturerÕs instructions;

$\bullet$		Discrepancy between the system design and the actual installation.
\smallskip

The summary of the T\"UV-Hessen 2011 final report indicated that the LPS were in   good condition, and the inspections could be repeated the following year on the 29 remaining sites. In 2012, inspections were performed the final time in the same manner as the previous years and a final report was generated by T\"UV-Hessen for all 29 sites. Out of these 29 sites, 28 site visits were completed because one site had a change in building management preventing the inspection from being completed.


\subsection{Strikes, flashes and Lightning Counters}
As explained in the [IEC 62305-1:2010] standard, a lightning flash is an electrical discharge of atmospheric origin between cloud and earth consisting of one or more strokes. According to this same standard, a single flash typically has between 3 and 4 strokes with each stroke having a different current waveform. The waveforms may be similar to that of an impulse (typical of first strokes and subsequent strokes) or to that of a continuous current (typical of a long stroke). Annex A of [IEC 62305-1:2010] introduces more details regarding the parameters and current waveforms of lightning flashes. 
 
A lightning counter such as the one installed on the LPS surveyed during this study counts multiple strokes for each single flash as long as the amplitude of a single stroke exceeds the minimum sensitivity of the counter and the response time of the counter is less than the interval time between subsequent strokes. In other words, the counter may not increment if the magnitude of the current stroke is small or if multiple strokes occur so rapidly that the counter cannot respond quickly enough. Thus strikes counted by the LPS must be corrected for multiple strike counts per flash. 
 
In [D'Alessandro and Petrov 2006], the authors addressed this counting issue by using the results of earlier publications [D'Alessandro and Darveniza 2001] in which a set of Monte Carlo simulations had been performed to determine a strike to flash ratio, or counting factor that was conservative, namely 2.5. This factor was only applied to counter readings where, from year-to-year, the reading had been increased by more than 1. When the counter reading was increased by only one on a year-to-year basis the counting factor was not applied. Despite the views of D'Alessandro and Darveniza this does not seem logical because an LPS counter record of 2 strokes would be equivalent to 0.8 flashes and a counter record of 1 stroke would be equivalent to 1 flash.  
 
Rakov, Uman, and Thottappillil (1996), however, point out that an unalterable path to ground for a given flash requires at least 4 consecutive strokes. Despite the fact that the same type of counter used by D'Alessandro and Petrov (2006) was used in this study, the Rakov, et al., results suggest the following  improved counting factor algorithm, based on electric field and TV observations,  for converting an LPS counter record of strikes to flashes: 
 
  (i) If  $\#$  Strokes counted by the LPS $\le$  3,   then  $\#$ Flashes =   $\#$  Strokes ;

   (ii) If   $\#$  Strokes counted by the LPS $\ge$  4,   then $\#$  Flashes =   $\#$ Strokes/3.5.

This rationale for converting strokes to flashes resolves an issue raised in [Abidin and Ibrahim 2002] relative to lightning stroke counter readings. As they point out these devices can give exceedingly high counts, some as high as 30 strokes in a single year. From the data used in our analysis, there was one incidence of 32 strokes in a single year, but when  the above indicated factor is taken into account  to estimate the number of flashes, the result is closer to 9 flashes per year. In the data available for our analysis all the other flashes per year ranged from zero to three. 

\subsection{Review of T\"UV-Hessen collected dataset}
An in-depth study of the data collected by T\"UV was undertaken using the individual inspection reports for each building as well as numerous photos. Based on the analysis, it was determined that five sites out of the 29 would be eliminated from the study due to issues of invalid or erroneous data collection.

For three out of the five sites, a zero percent efficiency was calculated. These data were rejected  because it conflicted with the expected number of strikes and the historical data from previous study. At another site, the lightning event counter went missing after the 2011 inspection.

Bypasses, or evidence of  lightning  damage  on  a  building,  were  identified  at  three  sites.  This is not surprising, considering that no lightning protection system offers 100$\%$ efficiency. On two of the three sites, one instance of damage to the building was identified by T\"UV. On the other building three bypasses were identified by T\"UV, all of them appearing within a one year span. The data suggest that this is a result of a single stroke having multiple attachment points to the structure. A careful reading of [D'Alessandro and Petrov  2006] and [Kong et al.  2009] indicates that multiple-ground terminations strokes (MGTS) are a very common occurrence. Taking this into account and considering the number of expected strikes on the structure, it was found that the three single bypasses should be counted as one single bypass.

These results must be contrasted with [Abidin and Ibrahim 2002]. They reported that more than 80$\%$ of the buildings in Kuala Lumpur on which un-conventional air terminals were installed had at least one lightning strike damage feature on them. But in our data the independent inspection by T\"UV indicated three damaged buildings out of a total of 24, i.e. 12.5$\%$ of the buildings were damaged. There is a statistically significant difference between the air terminal results of [Abidin and Ibrahim 2002] and those related to the 24 buildings in our study (P-value $<$ 0.001). In a review of the eleven photos in their report, which point out lightning damage, all but one appear to have a single ESE air terminal rather than an optimal placement of CVM air terminals for coverage as shown in blue in Figure 2.1. This might explain the significant difference between the two sets of results.


\subsection{Summary of final dataset}

In general, a valid statistical analysis requires at least 20 data points. However, it is impractical to collect statistically relevant data from a single structure because an exposure period of 30 years or more would be required. For this reason, and considering the stochastic nature of lightning, a successful field experiment must collect data from many structures. As shown in [D'Alessandro and Petrov 2006], even the analysis of sub-groupings of data is inappropriate. Hence, to make a comparison of the observed interception efficiency with the theoretical or estimated value, we need to use the entire data set.

In this study, many buildings are considered. Therefore, statistically relevant data were collected over a 3 year period, and all of the strike data from the year 2010 to 2011, and from 2011 to 2012, were combined to allow the analysis to be made of yearly data with statistical validity. The inherent statistical assumption in this approach is that all the data from buildings belong to the same population and that no buildings have features which differentiate them significantly from the rest of the buildings except height, which is accounted for in the various empirical forms of the equivalent exposure area, $A_{eq}$, discussed in Subsection 4.2. It was shown by [Petrov and D'Alessandro  2002], and [D'Alessandro and Petrov  2006], that these uncertainties are less than the fluctuations observed in normal lightning processes.

 \bigskip

 {\it Table 1:  Summary of data collected by T\"UV-Hessen used for statistical analysis}
  
 \vskip-.5cm
 
 \begin{table}[h!t!b!p!]
\vskip 5mm
        \centerline {
\begin{tabular}{|l|c|}
\hline 
Number of sites& 24\\
\hline  
Weighted average height of buildings, $\bar h_{weighted}$ & 70.1 meters \\
\hline
Total exposure time, $t_{total}$&37 years  \\
\hline
Average exposure times, $\bar t_{total}$&1.54 years  \\
\hline
Sum of individual number of flashes, $F_{observed}$&29.3  \\
\hline
Sum of individual number of bypasses, $B_{observed}$&3  \\
\hline
Sum of individual number of events, $\sum N_{d-observed}$&32.3  \\
\hline
Average number of events per year, $\bar N_{d-observed}$&0.873 \\
\hline
\end{tabular}
}
\end{table}

Table 1 shows a summary of the data collected by T\"UV-Hessen following the review process explained in the previous two sections.  The comparison of measured efficiency and theoretical efficiency based on the CVM analysis in Section 5 relies  on these results.


\section{Fractional Poisson Process  Model  for \\Predicting the Average Strikes per Year}
 
\subsection{Analysis of the lightning strike distribution}
In order to analyze the collected data in terms of comparison with the Eriksson attractive radius model, or in terms of lightning protection interception efficiency, it is necessary to check whether the frequency of our collected data matches a Poisson distribution utilized in  [D'Alessandro and Petrov  2006], who, employing  the Kolmogorov-Smirnov test,   argued  using their data that lightning flashes are randomly occurring rare events following that model.  However, applying this distribution model means that the assumption are (see, e.g., [Billingsley 1986]):

(i)	Lightning flashes occurring in non-overlapping intervals of time are statistically independent.

(ii)	The probability of the number of lightning flashes in a given interval of time depends on the length of the time interval.

(iii)	The probability of a single lightning flash in an infinitesimal time interval $dt$ is of the order   $\lambda\, dt$ where $\lambda $ is a positive constant.
	
(iv) The probability of more than a single lightning flash in an infinitesimal interval is zero.

With these postulates, consider a random variable  $X$ representing the number of lightning flashes in a year in a location which can take on values in   the set of positive integers,
$$
\Omega = \{0, 1, 2, 3, ...	\}\eqno{(1)} 
$$ 
with the probability distribution
$$
{\rm Prob} [X=k \,| \,\lambda]=e^{-\lambda}\frac{\lambda^k}{ k!} .
\eqno{(2)}
$$                                                                          
The constant parameter $\lambda$ in this equation, also called point estimator, is equal to the expected number of flashes in a year.

 Table 2 shows  the cumulative frequencies (cumulative distribution function, CDF) of observed flashes per year for the 37 years of data summarized in Table 1. 
Using the observed value for the average number of events per year from Table 1, 0.873, the Kolmogorov-Smirnov (K-S) maximum absolute difference between the observed and Poisson cumulative distribution function (CDF) is 0.204, which is significant  with a p-value of 0.10 (10$\%$). Thus  the hypothesis that the Poisson distribution represents the right model with these data cannot be rejected outright. 
 
 \newpage

 {\it Table 2: Cumulative frequency of the number  events per year from the 37 individual years}
 
 \vskip-.5cm
 
 \begin{table}[h!t!b!p!]
\vskip 5mm
        \centerline {
\begin{tabular}{|c|c|}
\hline 
$\#$ of Events per Year & Cumulative Relative Frequency\\
\hline  \hline
0 & 0.622 \\
\hline
1&0.838  \\
\hline
2& 0.946  \\
\hline
3 &0.973   \\
\hline
\end{tabular}
}
\end{table}

 An improved estimate of the mean number of events per year,  which was obtained by minimizing the maximum absolute difference between the Poisson and observed CDFs, turns out to be  is 0.563, a very different number than the empirical average.  The minimum K-S statistic in this case is 0.052, which indicates a low presumption against the Poissonian hypothesis. But even in this case the 95$\%$ confidence interval for the mean number of events per year is (0.321, 0.805), which does not include the observed value, 0.873.

Consequently, in this context we are proposing a fractional Poisson process as the lightning strike model with these data which can take into account the essential {\it burstiness} of the phenomenon. The model, developed by [Repin and Saichev  2000]  and  [Cahoy, Uchaikin, and Woyczynski   2010], has two parameters, $\lambda$, and $\nu$,  with $\lambda>0$, and 0$<\nu<1$.  The case $\nu=1$ corresponds to the standard Poisson process.  If $X$ is a fractional Poisson random variable representing the number of lightning flashes in a year in a location then, according to the fractional Poisson model, its probability distribution, the mean, and the variance are given by the following formulas: 
 $$
 {\rm Prob}\bigl  [X=n\,|\, \lambda,\nu\bigr]=
 \frac{\lambda^n}{n!}\sum_{k=0}^\infty \frac{(k+n)!}{k!} \cdot\frac{(-\lambda)^k}{\Gamma(\nu(k+n)+1)},
 \eqno{(3)}
 $$
 $$
 {\rm Mean} [X]=\frac{\lambda}{\Gamma(\nu+1)},
 \eqno{(4)}
 $$
  $$
   {\rm Variance} [X]= 
 \frac{\lambda}{\Gamma(\nu+1)} \left\{1+\frac{\lambda}{\Gamma(\nu+1)}\left[
   \nu 2^{1-2\nu} \frac{\Gamma(\nu)\Gamma(0.5)}{\Gamma(\nu+0.5)}-1\right]\right\}.
   \eqno{(5)}
   $$
   Here, $\Gamma(\nu)$ stands for the standard Euler Gamma function.  More   detailed discussion  about the rationale for our choice of the fractional Poisson model is included below in   Remark 1.

Using the fractional Poisson distribution, an estimate of the ${\rm Mean} [X]$  is 0.637, which was obtained by minimizing the maximum absolute difference between the fractional Poisson and observed CDFs,  resulting in a value of 0.633 for $\lambda$  and 0.030 for $\nu$ . The minimum K-S statistic  has now improved dramatically to  0.011 (from .204 for the standard Poisson model).  But more importantly, the mean number of events per year shown in Table 1, 0.873, is within 1.28 of the  standard error of the optimum mean, 0.637, the corresponding confidence interval being (0.383, 0.889).

Consequently, there is no presumption against using the fractional Poisson model in our case as the model also improves the fit with the observed CDF using the current data.

   \medskip{\bf Remark 1.} The basic idea of the fractional Poisson process (fPp) $X(t),$ motivated by experimental data with long memory (such as some network traffic, neuronal firings, and other signals generated by complex systems), is to make the standard Poisson model more flexible by permitting non-exponential, heavy-tailed  distributions of the inter-strike  times; the standard Poisson process has light, exponential tails of the inter-arrival times distributions.   Importantly,  fPp offers the ability to accommodate  clumping (burstiness) in the set of  lighting strike times (i.e., the set of jump points of  $X(t)$),  the phenomenon which naturally occurs  in production of lightning strikes. Such burstiness {\it cannot} be acomodated within   the standard Poisson process model. This substantial difference is clearly seen in Figure 4.1   which was borrowed from [Cahoy, Uchaikin and Woyczynski 2010].

   \begin{figure}[ht]
\centering
\numberwithin{figure}{section}
\includegraphics[width=12cm]{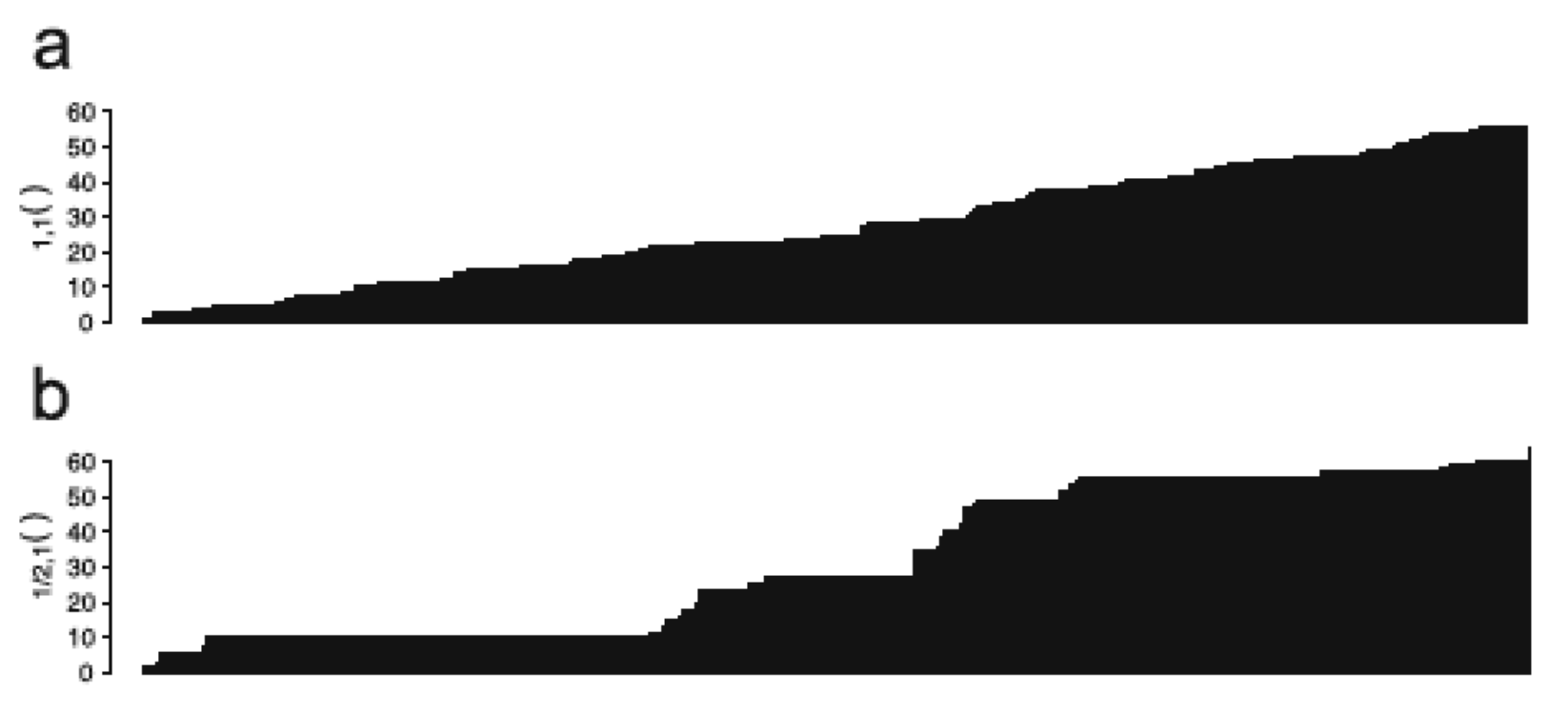}
\caption{Sample trajectories of: (a) standard Poisson process, (b) fPp with parameter $\nu$ = 1/2. The x-axis is time and the y-axis is the cumulative count of the fractional Poisson process.}
 \end{figure}

   However,  the price one has to pay for such flexibility is loss of the Markov property (i) from Subsection 4.1, a similar situation to that encountered in the case of certain anomalous diffusions studied by [Piryatinska, Saichev and Woyczynski   2005],  and [Woyczynski  2001]. Nevertheless, one can argue that the assumption (i) was too idealistic for the purpose of modeling the process $X(t)$ of cumulative counts of lightning strikes up to time $t$ not only because its inability to model the burstiness of the process, but also because of the well known long-range dependencies in the global (and local) weather patterns.    
   To partly replace this loss Markovianness one demands some scaling properties of the inter-strike  times' distributions which makes other tools such as the  fractional calculus available.  Thus, the probability distribution of the cumulative count $X(t)$ of lightning strikes by time $t$ is here defined as follows: 
    $$
 {\rm Prob}\bigl [X(t)=n\,|\, \lambda,\nu\bigr]=
 \frac{(\lambda t^\nu)^n}{n!}\sum_{k=0}^\infty \frac{(k+n)!}{k!} \cdot\frac{(-\lambda t^\nu)^k}{\Gamma(\nu(k+n)+1)}.
 \eqno{(6)}
 $$
 
For a detailed analysis of this process and the rigorous estimation procedures for its parameters, $\nu$ and $\lambda$, used in this paper, see [Cahoy, Uchaikin and Woyczynski 2010].  Also, note that fractal ideas have been applied in the lightning strikes context before,  but mainly to study the geometric fractal nature of the lightning  paths  themselves rather than in investigation of the temporal structure of the progression of lightning strikes as we are proposing in this paper.


 \subsection{Attractive radius calculations}
 
The number of expected strikes to the structures or events, $N_d$, also called number of dangerous events in [IEC 62305-2:2010], is determined using the equivalent exposure area, $A_{eq}$, and attractive radius, $R_a$, which was extensively documented in [Petrov and D'Alessandro 2002] using a field validation method, and is also included in [IEEE998-2012]. This concept is applied to revisit the number of dangerous events equation from Annex A of [IEC 62305-2:2010].

   \begin{figure}[ht]
\centering
\numberwithin{figure}{section}
\includegraphics[width=7cm]{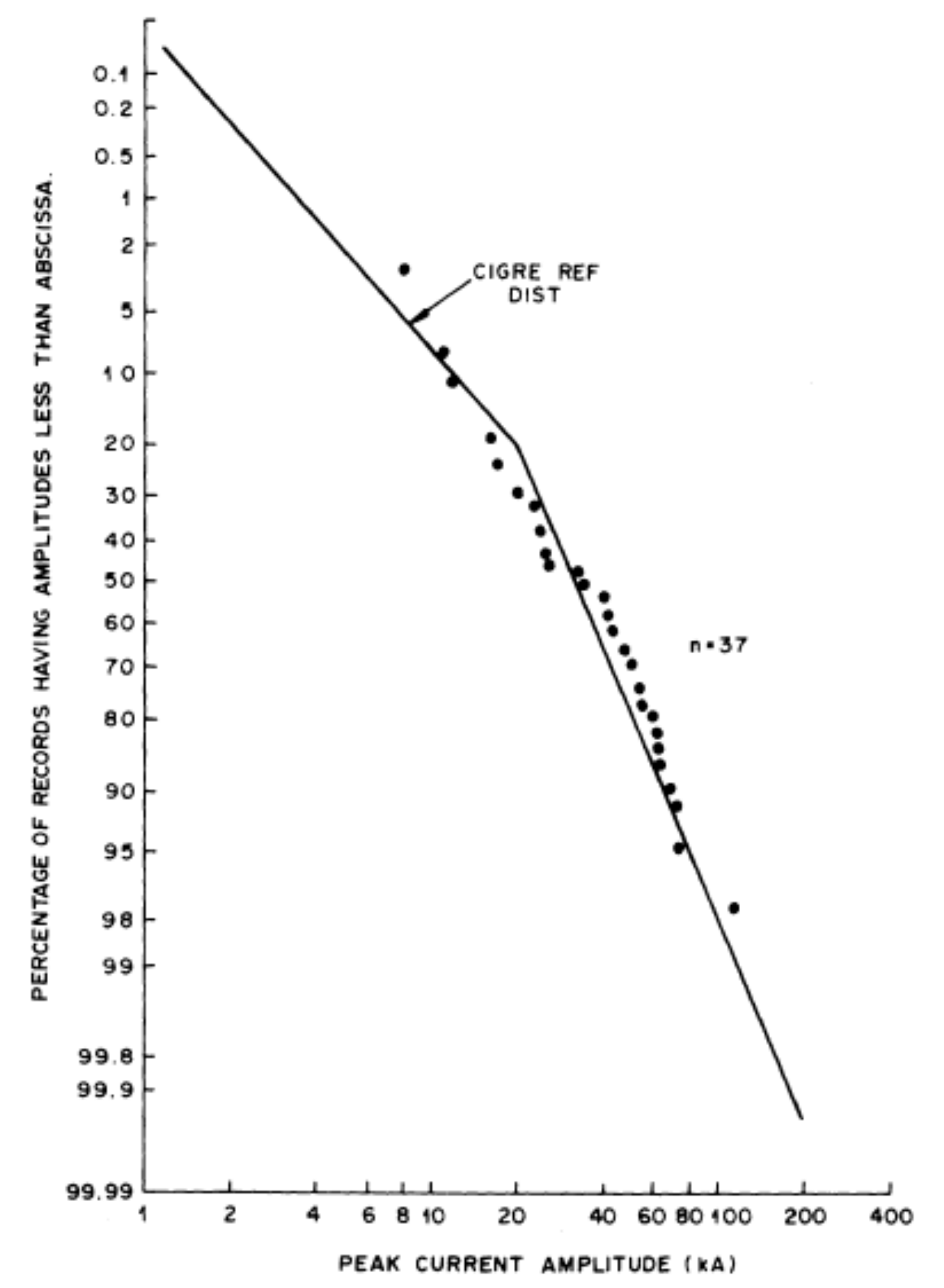}
\caption{Distribution of lightning peak currents amplitudes recorded during direct strikes over six years.}
 \end{figure}
 
 The number of strikes, or events,  to a structure, $N_d$, is determined by the following equation.
$$
N_d=N_gA_{eq}C_d10^{-6}.
\eqno{(7)}
$$
Here $A_{eq}$ is defined by the following integral, which is a probability-weighted average attractive area.
$$
A_{eq}=\pi\int_0^\infty R_a^2(i,h)f(i)\,di.
\eqno{(8)}
$$
In equation (8), $f(i)$ is the probability density function (PDF) of the peak current amplitude. 
In [Anderson and Eriksson  2006] it was  assumed that  the PDF to be log-normal-based despite the fact that [Eriksson  1980] pointed out that the peak current amplitude distributions of upward and downward flashes recorded on tall structures could involve mixtures of two sample distributions. In fact, Figure 7 in  [Eriksson  1980],  and Figure 2, p. 869, in [Eriksson 1987], reproduced here as  Fig. 4.2, clearly reveals that the PDF of the peak current density is bi-variate or a mixture of two distributions. 

In response to Eriksson's suggestion to study this bi-variate nature of the peak current density, we have analyzed the shape of the CDF in his Figure 7 and found it to be a mixture of two log-normal distributions, with 20$\%$ from the lower kA or shielding currents, and 80$\%$ from the higher kA back-flash currents . The analysis of this mixture is based on the fact that the tail of each distribution is minimally contaminated by the other mixture distribution as indicated by the linearity of the data when the CDF is plotted as a log-normal probability graph. The points on the linear portion of the CDF in each tail, adjusted for the percentage in the mixture as shown in the following equations, produce estimates of the mean and standard deviation of each component in the mixture.
 $$
Adjusted\;CDF_{shielding}=\frac{CDF_{lower\;tail}}{0.20}
\eqno{(9)}
$$ 
 $$
Adjusted\;CDF_{back\;flash}=\frac{CDF_{upper\;tail}-0.20}{0.80}
\eqno{(10)}
$$
From Eriksson's Figure 7 the lower kA shielding current distribution is log-normal with a mean (natural log) of 2.48, and a standard deviation (natural log) of 0.91. The higher kA back-flash current distribution is log-normal with a mean (natural log) of 3.66, and a standard deviation (natural log) of 0.53. From this bi-variate mixture of two log-normal distributions, the mean value of the peak current amplitude is 39.7 kA, which agrees with the approximately 40 kA figure in [Anderson and Eriksson  2006], but it is significantly larger than the median peak current value, of 33kA as suggested in [Yahaya and Zain  2000].

Three empirical models were presented in  [D'Alessandro and Petrov 2006] for the attractive radius, $R_a$:
$$
R_a=0.84i_p^{0.74} h^{0.6},
\eqno{(11)}
$$
see, Eriksson (1987 a,b),	
$$
R_a=0.56[(h+15) i_p ]^{2/3}   
\eqno{(12)}
$$ 	                                  
see, [Petrov and  Waters 1995, Petrov et al 2000], and 
$$
     R_a=25.9h^{0.48}
     \eqno{(13)},
     $$                                                                    
see, [Rizk  1994 a,b].     In these equations, $H$ is the height of the structure in meters and $i_p$ is the peak current in kA.

Using the mixture of log-normal distributions, $A_{eq}$ is expressed as a function of the the height, $h$,  of the structure as follows:
     $$
A_{eq}=594\cdot h^{0.6},
\eqno{(14)}
$$
for the  Eriksson (1987 a,b) empirical model (11),	
$$
A_{eq}=295 (h+15)^{2/3},
\eqno{(15)}
$$ 	                                  
for the Petrov and Waters (1995), Petrov et al (2000),  model (12) 
 $$
A_{eq}= 2107\cdot h^{0.96}
     \eqno{(16)}
     $$                                                                    
for the [Rizk  1994 a,b] model (13).

The constant, $C_d$, appearing the the formula (7) predicting the number of strikes to a structure, is the location factor of the structure based on Table A.1 of [IEC 62305-2:2010] which has been reproduced in Table 3. This location factor was determined by carefully looking at the pictures sent by T\"UV-Hessen and picking the closest conservative value. Sixteen of the 24 buildings in our data had Cd values of 0.25 and the remainder were 0.50.
\bigskip

{\it Table 3. Structure location factor, $C_d$}

  \begin{table}[h!t!b!p!]
\vskip 5mm
        \centerline {
\begin{tabular}{|l|c|}
\hline 
{\bf Relative Location} & $C_d$\\
\hline  \hline
Structure surrounded by higher objects & 0.25 \\
\hline
Structure surrounded by objects of the same height or smaller&0.50  \\
\hline
Isolated structure: no other objects in the vicinity& 1.00 \\
\hline
Isolated structure on a hilltop or a knoll &2.00  \\
\hline
\end{tabular}
}
\end{table}


\subsection{ Lightning activity in the Klang Valley \\during the 2010-2012 period}
The weighted average ground flash density $\bar N_g$ has been estimated based on official data obtained from the research division of the Malaysian national utility Tenaga Nasional Berhad (TNB). TNB Research Sdn Bhd operates the lightning detection network across Malaysia, and the results shown in Table 4 were obtained for the period ranging from 2010 to 2012, which is  also  covered in this study.

\bigskip

{\it Table 4: Average ground flash density for various locations around the Klang Valley area}

 \begin{table}[h!t!b!p!]
\vskip 5mm
        \centerline {
\begin{tabular}{|l|c|}
\hline 
{\bf Area} &{\bf  $N_g$ in flashes/km$^2$/year}\\
\hline  \hline
KL Sentral & 20 to 28 \\
\hline
Shah Alam, Selangor&24 to 32  \\
\hline
Subang Jaya, Selangor& 28 to 32  \\
\hline
Putrajaya&20 to 24   \\
\hline
\end{tabular}
}
\end{table}

\bigskip

The weighted average values for the minimum, midpoint, and maximum ground flash density, are 22.2, 25.7, and 29.2 flashes per km$^2$, per year, respectively for the 37 combined exposure years.


\subsection{Result of the comparison between the number of \\expected strikes and the collected data}
 The theoretical probability-weighted number of flashes per year to the structures studied based on Equation (7) was calculated using the Petrov and  WatersÕ, Eriksson, and Rizk empirical equations for $R_a$ depending on whether the low, midpoint, or the high estimate of the ground flash density, $N_g$, from Table 4 was used. The results are shown in Table 5.

From Table 1, the observed value of , $\bar N_{d-observed}$, 0.873, can be compared with the theoretical probability-weighted number of flashes per year at the midpoint value of $N_g$, 0.861. Table 5 illustrates that Eriksson's  empirical model for $R_a$ is more consistent with the T\"UV data than either the Petrov and Waters, or the Rizk models, and that the mixture of two log normal distributions improves upon the theoretical predictions using the univariate normal distribution for the peak current amplitude.
\medskip

{\it Table 5: Summary of average number of strikes per year based on different PDF for peak current amplitude and different values for average ground flash density}

  \begin{table}[h!t!b!p!]
\vskip 5mm
        \centerline {
\begin{tabular}{|l|c|c|c|c|}
\hline 
{ Theoretical $N_d$}& {  Mixture of Two} &{  Mixture of Two} &{  Mixture of Two}&{ Univariate Log}\\
\hline
 & {  Log Normals} &{  Log Normals} &{  Log Normals}&{   Normal}\\
\hline  \hline
{  Avg $\#$ Strikes/Year}&{  Petrov $\&$ Waters}&{  Eriksson}&{  Rizk}&  {  Eriksson}\\
\hline
{  $N_g$ Low}& 0.860&0.748&0.923&0.806 \\
\hline
{ $N_g$ Midpoint} &0.991&0.861&1.062& 0.928  \\
\hline
{ $N_g$ High} &1.121&0.975&1.200& 1.051  \\
\hline
$\%$ Error:  $N_{d-N_{g Midpoint}}$&-13.6 $\%$&1.3 $\%$&-21.9 $\%$&  -6.4 $\%$ \\
 versus $N_{g\, Observed}$&&&&  \\
 \hline
\end{tabular}
}
\end{table}


\subsection{ Comparison between the theoretical and actual lightning protection system interception efficiency}
Following the positive results shown above based on the use of the mixture of two log-normal distributions for the peak current amplitude and the fractional Poisson distribution for the number of strikes per year, it is now possible to assess whether the actual interception efficiency, $E_{observed}$ of the CVM-based and optimally positioned lightning protection systems installed on buildings surveyed by T\"UV  corresponds to the theoretical efficiency $E_{theoretical}$. Of the 24 buildings comprising the final data set, 2 are protected by a lightning protection system with efficiency 97$\%$, 20 with a lightning protection system with efficiency 91$\%$, and 2 with a lightning protection system with efficiency 84$\%$. This yields an average theoretical efficiency $E_{theoretical}$ of 90.9$\%$. On the other hand, $E_{observed }$ is determined from Table 1 by equation (17). 
$$
E_{observed}=\frac{F_{observed}}{N_{observed}	}.	
\eqno(17)
$$
 That is, the actual interception efficiency, $E_{observed}$, of the CVM-based lightning protection systems surveyed by    T\"UV  is 90.7$\%$, which is in very close agreement with the $E_{theoretical}$, the theoretical lightning protection efficiency. The error between $E_{observed}$ and $E_{theoretical}$ is minor and understandable. As a side comment, it should be noted that if the 2.5 counting factor from Subsection 4.2 had been applied across all sites regardless of whether the number of strikes counted by the LPS was 1 or more, the observed efficiency would have been 90.3$\%$, while if the counting factor has been   applied using  [D'Alessandro and Petrov 2006], the efficiency would have been 91.0$\%$.


\section{ Conclusions}
The purpose of this paper was to resolve a question posed by [Abidin and Ibrahim 2004] regarding the effectiveness of a LPS based on the CVM air terminals optimally positioned on buildings to include the competing features of the building. Toward this end, two research objectives were addressed. The first was to use lightning strike data from 24 buildings in Kuala Lumpur, Malaysia, that were gathered over a two year period to estimate the extent of protection provided by CVM air terminals optimally positioned on buildings. The second objective was to investigate the validity of the various underlying semi-empirical theories for CVM air terminals.

	Relative to the first objective, our statistical analysis of the strike data recorded on the 24 buildings for 37 combined years of service indicated that the observed efficiency is 90.7$\%$ with CVM air terminals optimally positioned on buildings in Kuala Lumpur. These data significantly contradict (P-value $<$ 0.001) the 80$\%$ reported damage rate in Kuala Lumpur due to unconventional air terminals [Abidin and Ibrahim 2004].    
	
	Relative to the second objective, following the approach of [Petrov and D'Alessandro 2002],  and [D'Alessandro and Petrov 2006] , we have demonstrated that the fractional Poisson distribution improves the fit between actual and predicted cumulative distribution functions of lightning strike data that were recorded in Kuala Lumpur. This further validates the CVM and Erikson's attractive radius model on which the CVM is based. As expected, no lightning protection system offers 100$\%$ protection. But the observed 90.7$\%$ interception efficiency with LPS components optimally placed according to the CVM coverage criteria illustrated in Figure 2.1 is in accordance with the theoretical efficiency. 
	
	 All of the CVM models we researched are semi- empirical and, hence, semi-theoretical, i.e. dependent upon estimates from other data to adjust for peak current amplitudes, structural features, and ground flash densities. As such, these models may not account for all physical phenomena. But as George Box, chairman of the Department of Statistics at the University of Wisconsin,  frequently remarked to his students, ``All models are incorrect, but some are useful''. Our comparison between the analysis of independently observed lightning strike data and CVM models indicates that the Eriksson model agrees with our data and, therefore, can be useful. 
	
	It is important to note that the theoretical efficiency of each site is based on a minimal peak current for each protection level, which is equivalent to the parameters listed [IEC 62305-1:2010]. It has been observed that the differences between the theoretical and observed efficiency is less than 0.5$\%$ . For comparison, the differences between observed and theoretical values in the initial study is 1.5$\%$  [D'Alessandro and Petrov  2006].
	
	This variation can be explained by the stochastic nature of lightning or, alternatively, by the fact that in this study LPS were analyzed that utilized a newer generation of CVM air terminals combined with optimal positioning of air terminals. These terminals have a blunt tip and optimized dome shape resulting from a careful application of the extensive research of [Moore et Al.  2003] and [D'Alessandro et al 2003].
	
	It has also been shown that the Eriksson attractive radius concept can be applied when using risk assessment calculations per IEC 62305-2, as the null hypothesis testing has demonstrated that the number of events predicted by the theoretical model is in line with the field data collected.

\bigskip

{\bf Acknowledgements:} 
Our research consisted of an independent statistical study of data gathered by T\"UV-Hessen from measurements using the ERICO$ ^{\footnotesize\textregistered}$ SYSTEM 3000 lightning protection system, as well as TUV HessenÕs annual structural inspections, in order to validate and expand earlier studies by A. J. Eriksson, R. B. Anderson, F. D'Alessandro, and N. I. Petrov. We are grateful to T\"UV-Hessen for providing these data and ERICO for sponsoring our research.   We would also like to thank the anonymous referees for their help   to improve the original version of this paper.

\section{References}
.
  
       \smallskip 

[Abidin and Ibrahim 2002] H. Z. Abidin and R. Ibrahim, Conventional and Un-conventional Lightning Air Terminals: An Overview, Forum on Lightning Protection, Hilton Petaling Jaya, January 2004 

       \smallskip 
    [Anderson and Eriksson 1980] R.B. Anderson and A.J. Eriksson, Lightning parameters for engineering applications,  Electra  69, 65-102, 1980.
    
      \smallskip 
    [Billingsley 1986] P. Billingsley, Probability and Measure, Wiley and Sons, New York, 1986.
  
 \smallskip            
   [Cahoy, Uchaikin, and Woyczynski 2010]   D.O.Cahoy, V.V. Uchaikin and W.A. Woyczynski, W. A.,   \textit{Parameter estimation for fractional Poisson processes}, Journal of Statistical Planning and Inference 140 (2010), 3106-3120.   
   
     \smallskip     
[D'Alessandro 2003] F. D'Alessandro, The use of  field  intensiÞcation factors in calculations for lightning protection of structures J. Electrostat. 58, 17-43, 2003.

 \smallskip      
[D'Alessandro and Darveniza 2001] F. D'Alessandro,   and  M. Darveniza,  Field validation of an air terminal model, Proc. Sixth Int. Symp. on Lightening Protection, Santos, Brazil pp. 234239, 2001.  
     
  \smallskip      
[D'Alessandro and Petrov 2006] N.I. Petrov  and  F. D'Alessandro, Field study on the interception efficiency of lightning protection systems and comparison with models, Proc. R. Soc. Lond. A 462, 1365- 1386, 2006.

 \smallskip      
[D'Alessandro and Gumley 2001] F. D'Alessandro,   and  J.R. Gumley,   A  collection volume method  for the placement of air terminals for the protection of structures against lightning, J. Electrostat. 50, 279-302, 2001.

 \smallskip      
[D'Alessandro et Al. 2003] F. D'Alessandro, C. J. Kossmann, A. S. Gaivoronsky and  A. G. Ovsyannikov, Experimental Study of Lightning Rods Using Long Sparks in Air, IEEE Transac. on Dielectrics and Electrical Insulation Vol. 11 No. 4, 638-649, 2003.

  \smallskip     
[D'Alessandro 2003] F. DÕAlessandro,  The use of field  intensiÞcation factors in calculations for lightning protection of structures J. Electrostat. 58, 17-43, 2003.

\smallskip

     [ERICO 2009] ERICO$ ^{\footnotesize\textregistered}$ , ERITECH SYSTEM 3000 Installation, Operation and Maintenance Manual, 2009. 
     
       \smallskip     
    [ Eriksson 1979] A.J. Eriksson, The lightning ground flash: an engineering study, Ph.D. Dissertation, University of Natal, Pretoria, South Africa, 280 pp., 1979

    \smallskip 
    [Eriksson 1987 a,b]  A.J. Eriksson, The incidence of lightning strikes to power lines, IEEE Transaction on Power delivery, PWRD-2, 859-870, 1987.
     
 \smallskip      
[IEEE998-2012] IEEE Power and Energy Society, IEEE998-2012-IEEE Guide for Direct Lightning Stroke Shielding of Substations, 2012
     
 \smallskip      
[IEC 62305-1:2010]  IEC, Protection against lightning   Part 1:  General principles, Ed 2.0, 2010
     
\smallskip       
[IEC 62305-2:2010] IEC, Protection against lightning  Part 2: Risk Assessment, Ed 2.0, 2010

\smallskip
[Klyatskin and Woyczynski 1997] V.I. Klyatskin and W.A. Woyczynski, Dynamical and statistical characteristics of geophysical fields and waves and related boundary-value problems, in Stochastic Models in Geosystems, S.A. Molchanov and W.A. Woyczynski, Eds., Springer-Verlag, New York, 1997, pp. 171-208.
     
\smallskip       
[Kong et Al. 2009] X.Z. Kong, X.S. Qie, Y. Zhao and  T. Zhang,  Characteristics  of  negative  lightning flashes presenting multiple-ground terminations on a millisecond-scale, J. Atmospheric Research 91, 381-386, 2009.

\smallskip 
[Molchanov and Woyczynski 1997] S.A. Molchanov and W.A. Woyczynski, Eds., Stochastic Models in Geosystems,
Springer-Verlag, New York, 1997, 492 pp.
     
 \smallskip      
[Moore et Al. 2003] C.B. Moore, G.D. Aulich, and  W. Rison, The case for using blunt-tipped lightning rods as strike receptors, J. Appl. Meteorology 42, 984-993, 2003.

     \smallskip
 [Neuts 1993] M.F. Neuts (1993.) The Burstiness of Point Processes,  Commun. Statist.ÑStochastic Models, 9(3):445Ð466, 1993.

     \smallskip
[Petrov and Waters 1995] N.I. Petrov and R.T. Waters,      Determination of the striking distance of lightning to earthed structures, Proceedings of the Royal Society of London, Series A 450, 589--601, 1995.
     
        \smallskip
[Petrov, Petrova, and D'Alessandro 2003]  N.I. Petrov, G.N. Petrova  and F. D'Alessandro,
Quantification of the probability of lightning strikes to structures using a fractal approach, IEEE Transaction on Dielectrics and Electrical Insulation 10, 641-654

     \smallskip
[Petrov, Petrova, and Waters .2000]  N.I. Petrov, G.N. Petrova  and R.T. Waters,
Determination of attractive area and collection-volume  of earthed structures, Proceedings of the 25th ICLP, Rhodes, Greece, ppl 374-379, 2000.

       \smallskip
[Petrov and D'Alessandro 2002] N.I. Petrov  and  F. D'Alessandro, Assessment of protection system   positioning and models using observations of lightning strikes to structures, Proc. R. Soc. Lond. A 458, 723-742, 2002
     
    \smallskip
[Petrov and D'Alessandro 2002] N.I. Petrov and  F. D'Alessandro, A verification of lightning strike incidence as a Poisson process, Journal of Atmospheric and Solar-Terrestrial Physics 64, 1645-1650, 2002.

  \smallskip    
[Piryatinska,  Saichev , and Woyczynski 2005]  A. Piryatinska, A.I. Saichev, and W.A. Woyczynski,   Models of anomalous diffusion: The subdiffusive case,   \textit{Physica A: Statistical Physics}, \textbf{ 349} (2005), 375-424.

            \smallskip  
[Rakov et al. 1994] V.A. Rakov, M.A. Uman, and R. Thottappillil, Review of lightning properties from electric field and TV observations, Journal of Geophysical Research,  99,  D5 (1994),   10,745 -10,750.

      \smallskip  
[Repin and Saichev 2000] O. N. Repin and A. I. Saichev, Fractional Poisson Law, Radiophysics and Quantum
     Electronics 43, 9 (2000), pp. 738 - 741
      
      \smallskip
      
[Rizk 1994 a,b] F.A.M. Rizk, Modeling of lightning incidence to tall structures, Part II:Application, IEEE Transactions on Power delivery 9, 172-193, 1994.

  \smallskip        
[Woyczynski 2001]   W.A. Woyczynski,    \textit{L\'evy Processes in the physical sciences, } in {\it L\'evy processes}. Barndorff-Nielsen, O., Mikosch , T., and Resnick, S., Eds., Birkh{\"a}user -Boston, 2001. 
   
    \smallskip      
[Yahaya and Zain 2000] M.P. Yahaya,   and   S. A. F. S. Zain, Characteristics of cloud-to-ground lightning in Malaysia, Conf. Lightning Protection and  Earthing Systems, Kuala Lumpur 1-15, 2000.

\end{document}